\documentclass[a4paper,wide]{article}

\usepackage{times}
\usepackage{ifpdf}
\usepackage{cite}
\usepackage[pdftex]{graphicx}
\usepackage{array}
\usepackage{mdwmath}
\usepackage{mdwtab}
\usepackage{amssymb,latexsym}
\usepackage{stfloats}
\usepackage{amsmath,esint}
\usepackage{subfig}
\usepackage{booktabs}

\newcommand{\imaginary}{\mathrm{j}}

\usepackage{algcompatible}

\usepackage{algorithm}

\usepackage[font={footnotesize}]{caption}

\usepackage{balance}

\usepackage[utf8]{inputenc}

\usepackage[T1]{fontenc}

\usepackage[compatible]{algpseudocode}

\begin{document}

\title{
Critical Near-Field Impedance Matrices}
%
%
%
\author{   Peter Krämer\\
{University Freiburg}\\
Freiburg, Germany \\
kraemerp@informatik.uni-freiburg.de
\and
 {Christian Schindelhauer} \\
   {University of Freiburg, Germany} \\
   {schindel@informatik.uni-freiburg.de}
     }

\maketitle

	\begin{abstract} 
	We investigate the theoretical impedance equations for several near-field antenna positions.
	In the standard model one computes the currents at the antennas for given voltages using the impedance matrix of the antennas, which is only possible if the determinant of the impedance matrix is non-zero. We consider Hertzian group antennas, its relative corresponding impedance  and two approximations (mid and far) of it. For the approximations we  show that for many situations the determinant is zero.

We find three antenna configurations for three antennas, i.e., on a line, on a right triangle, and an isosceles triangle, which result in a zero determinant of the impedance for the far-field approximation. This means that with existing methods, one cannot determine the behavior of this antenna system. For the better mid approximation, we find a configuration of 15 triangular-positioned antennas resulting in a singular impedance matrix. 

Furthermore, we investigate $n\times n$ grid placed antennas in the more accurate Hertzian impedance model and find that for $d \approx 0.65$ 
 wavelengths of grid distance for $n=2, \ldots, 8$ the absolute value of the determinant of the corresponding impedance matrix decreases by an order of magnitude with each increased grid size.

\end{abstract}
\bigskip
\noindent
\bigskip

\centerline{{\bf Keywords}: Antenna theory, Dipole antennas, Linear Antenna arrays, Radiating Impedance, Hertzian model}
\section{Introduction}
We investigate radiation-coupled thin dipole antennas, which are fed into a circuit consisting of a generator, matching impedance and the actual antenna  \cite[427]{kark2004antennen}. 
Our considerations are limited to isotropic radiators and Hertzian dipoles to ensure a consistent solution approach.

The energy radiated by the antenna (Poynting vector) is used to determine the electric and magnetic vector field with its near and far field components. The radiation impedance of the individual antenna calculated from this, with the reactive near field and real far field (TEM-Wave), serves as the basis for determining an impedance matrix of an antenna array of mutually influencing individual antennas.

This directly results in requirements regarding the determinant of this impedance matrix and the associated solvability of the linear system of equations, since the determination of antenna correction currents requires the existence of an inverse impedance matrix. 

\section{Related Work}
As a basics, the paper is based on Schelkunoffs work \cite{schelkunoff1953antennas}. An radiated power approach, including near and fare ($kr \ll 1$ to $kr \gg 1$) field, we find at Balanis \cite{Balanis15}. The fields of the Hertzian dipole and in particular the reactant field were considered in the work of Schantz \cite{schantz2001electromagnetic}.  
A system theoretical approach with isotropic antennas in Uniform Linear Array ULA configuration is given in \cite{Tse_fundamentals_book}. 
We refer here also to the basic works of Miki and Antar \cite{mikki2011theory, mikki2011theoryb} on the subject of antenna near field of a single antenna. Vendik et al. develops the complex antenna impedance for mutually influencing antenna pairs based on the Kramer-König relation \cite{vendik2015novel}.
The impulse behavior of the energy propagation in the reactive near field of the antenna is described in the article by Valagiannopoulos and Al$\acute{u}$ \cite{valagiannopoulos2015role}. The near-field behavior of future very large antenna arrays described by Cui et. al. is of fundamental importance even for the new 6G generation of mobile communications \cite{cui2022near,cui2022channel}.

Yordanov et al. \cite{yordanov2009arrays} analyses  the impedance of individual antennas for the Hertzian-Dipole in a ULA arrangement is calculated. From this, they  determine the mutual impedance change caused by the presence of additional antennas. 
The concepts of multi port theory will be expanded in the follow-up works of Ivrala\v{c} et al. \cite{ivrlavc2010multiport} and Phang  et al. \cite{phang2018near}. This provides a basis for signal transmission in MIMO channels.
Finally, we refer here to the relevant work of Zuhrt \cite{zuhrt1953energieverhaltnisse} and Kark \cite{kark2004antennen}.

\section{Model and Notations}
Commonly the impact of antennas is classified into near field, Fresnel and Fraunhofer-zone, which include field components $\vec{e_\vartheta}, E_\vartheta $ with $1/r^3, 1/r^2 ,1/r$ reduction and $H_\varphi$ with $1/r^2, 1/r$ reduction.

Starting with the radiation impedance of a single antenna, an impedance model can be  derived as follows.
With the help of superposition, we calculate the sum of vector components of the electric field. In a further step, the radiation impedance of antenna arrangements is calculated from this in the form of a matrix. The sum of the load reactance and impedance of individual antenna is in the diagonal values of the matrix. 	

The radiated power is given by the Poynting vector existing of a $\vartheta$ and $ r$ component, where only the $E_\vartheta$ component makes a contribution to the far field \cite{Balanis15}.
\begin{equation}
	W = \frac{1}{2} (E \times H^*) = \frac{1}{2} (e_r E_r + e_\theta E_\theta) \times (e_\Phi H_\Phi^*)\label{Energy}
\end{equation}
%
\subsection{Hertzian Impedance}

We choose the Hertzian-Dipole as the radiation element in our analysis, since a complete  mathematical solution  is known. Using the radiated energy, we can now determine the intrinsic impedance of the Hertzian dipole and the mutual relative impedance of two dipoles for a unified linear array (ULA).

For the wavelength $\lambda$ we have the wave number $k = \frac{2\pi}{\lambda}$. 
First, we describe the real part $R_{nn}$ and imaginary part  $X_{nn}$ of the intrinsic impedance of the Hertzian dipole, which results from the radiated power.\cite{yordanov2009arrays}
\begin{equation}
	R_{nn}=Z_0 \left(\frac{2\pi}{3}\right)\: \left( \frac{\ell}{\lambda}\right) ^{2}
\end{equation}
\begin{equation}
	X_{nn}=Z_0 \left(\frac{2\pi}{3}\right)\: \left( \frac{\ell}{\lambda}\right) ^{2} \frac{1}{(kr)^3}
\end{equation}
where $Z_0 = 376.73 \ldots \Omega$ is the impedance of free space. For $r\rightarrow \infty$ the self impedance $Z_{nn} = R_{nn} + \imaginary X_{nn}$ will be completely dominated by the real part, since $X_{nn} \rightarrow 0$ and thus we will set $Z_{nn}= R_{nn}$.

	Due to the uniform radiator arrangement, the real part of the mutual impedance of two antennas is calculated using the $\Psi$ function as follows
\begin{equation}
	R_{mn}= R_{nn} \Psi (kd) \ , \hbox{where}
\end {equation}
\begin{equation}
	\Psi(x) = \frac{3}{2} \left( \frac{\sin x}{x}\:+ \frac{\cos x}{x^2}\:- \frac{\sin x}{x^3}\right)
\end{equation}
and  $x = k d = \frac{2\pi d}{\lambda}$. Here  $d$ is the distance of the two parallel antennas, perpendicularly placed on the same plane. Throughout this paper we consider all antennas in a two dimensional setting placed parallel perpendicurlarly on the same plane.

In the same way, we obtain the imaginary part by considering the $\Phi$ function.
\begin{equation}
	X_{mn}= R_{nn} \Phi(x)  \ , \hbox{where}
\end{equation}
\begin{equation}
	\Phi(x)=  \frac{3}{2} \left( \frac{\cos x}{x}\:- \frac{\sin x}{x^2}\:- \frac{\cos x}{x^3}\right) \ .
\end{equation}
So, the relative impedance can be computed as
\begin{equation}
	Z_{mn} = R_{nn} f(x)\ ,
\end{equation}
where
\begin{equation}
	f(x) = \Psi(x) + \imaginary \Phi(x)\quad  \mbox{and}
\end{equation}
\begin{equation}
	f(x) = \frac{3}{2} \left( \imaginary \frac{e^{-\imaginary x}}{x}+ \frac{e^{-\imaginary x}}{x^2} - \imaginary \frac{e^{- \imaginary x}}{x^3}\right)\ .
\end{equation}
Note that we denote the imaginary unit with  $\mathrm{j}$.
So, we obtain the complex function $f(x)$ which allows us to calculate the mutual impedance of two dipoles using $R_{nn}$ by substituting $x$ by $kd$ as follows
\begin{equation}
	Z_{mn} =  \frac{3}{2}\ R_{nn}\   e^{-\imaginary kd}\left(\frac{\imaginary}{kd}+ \frac{1}{(kd)^2}- \frac{\imaginary}{(kd)^3}\right)\ .
\end{equation}

\begin{figure}[h]
\begin{center}
	\includegraphics[width=0.7 \linewidth]{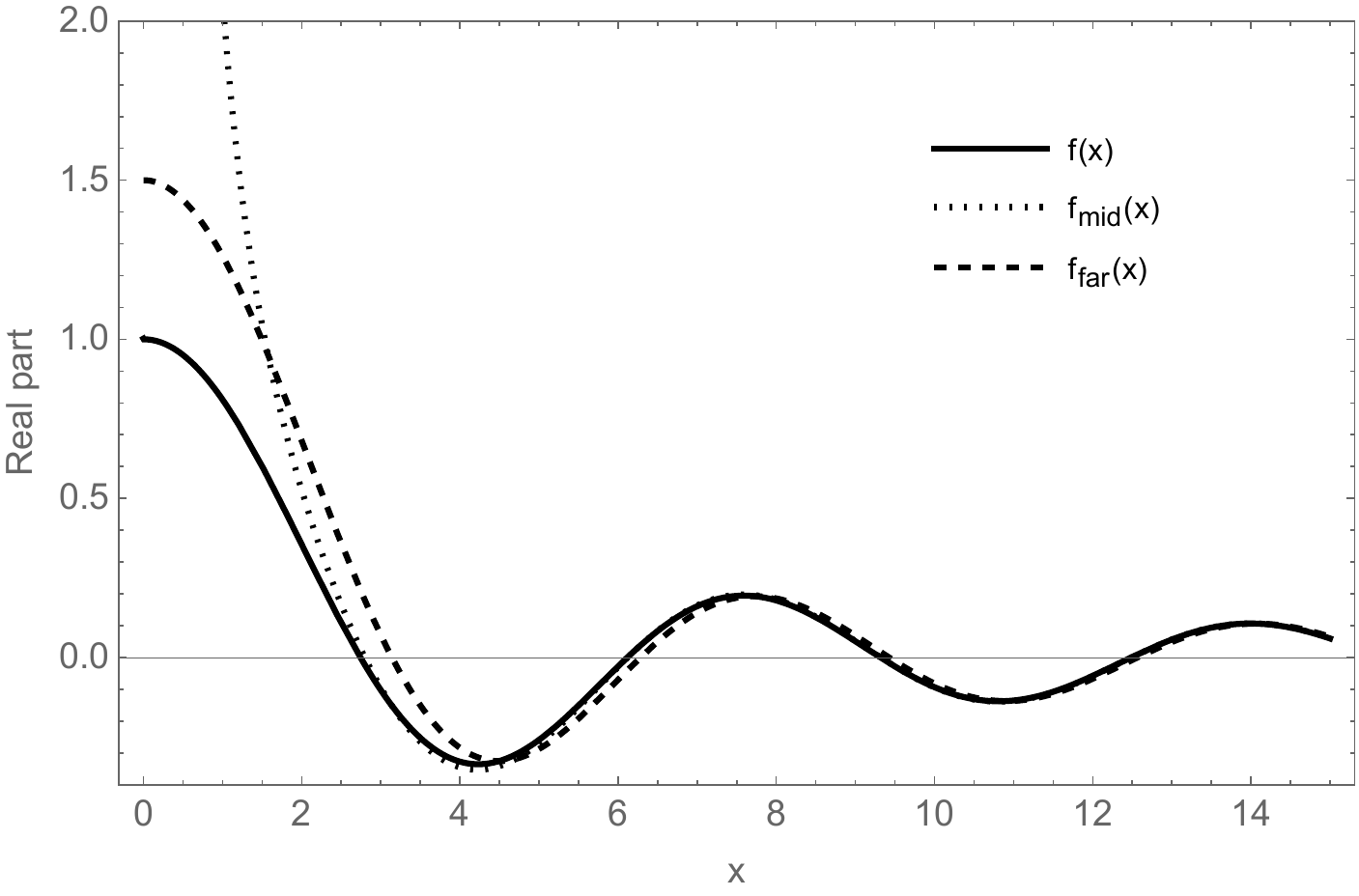}
	\caption{Real part of $f$ and its approximations $f_{\mbox{\scriptsize mid}}$  and $f_{\mbox{\scriptsize far}}$ \label{effs} }
	\end{center}
\end{figure}

\begin{figure}[h]
\begin{center}
		\includegraphics[width=0.7 \linewidth]{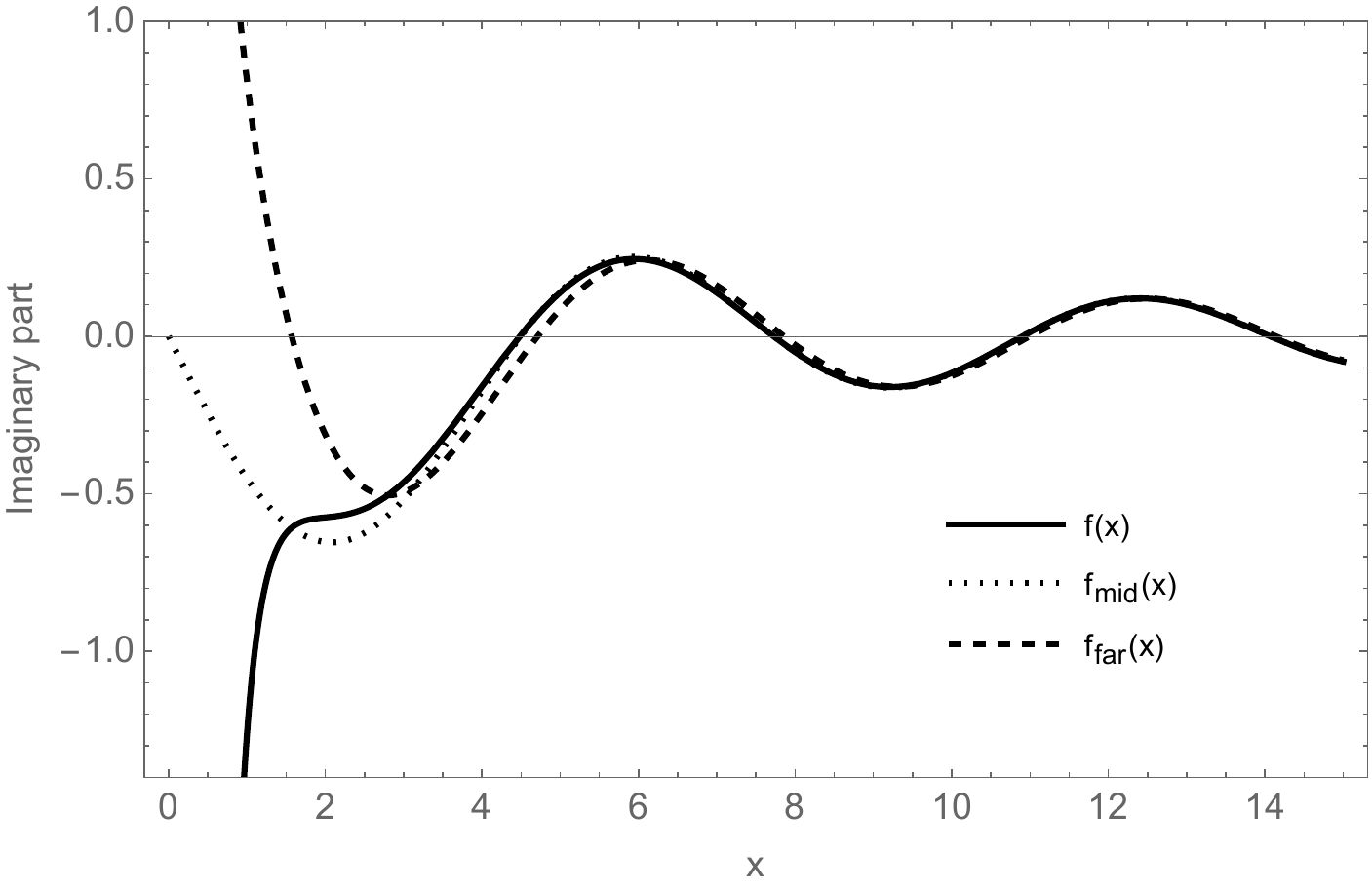}
	\caption{Imaginary part of $f$ and its approximations $f_{\mbox{\scriptsize mid}}$  and $f_{\mbox{\scriptsize far}}$ \label{effs} }
	\end{center}
\end{figure}
The individual curves greatly differ  for $x<2$, especially for imaginary part and cannot regarded as an  approximation there.
\subsection{Impedance Matrix}
\begin{figure}[h]
\centering
\includegraphics[width=0.7\linewidth]{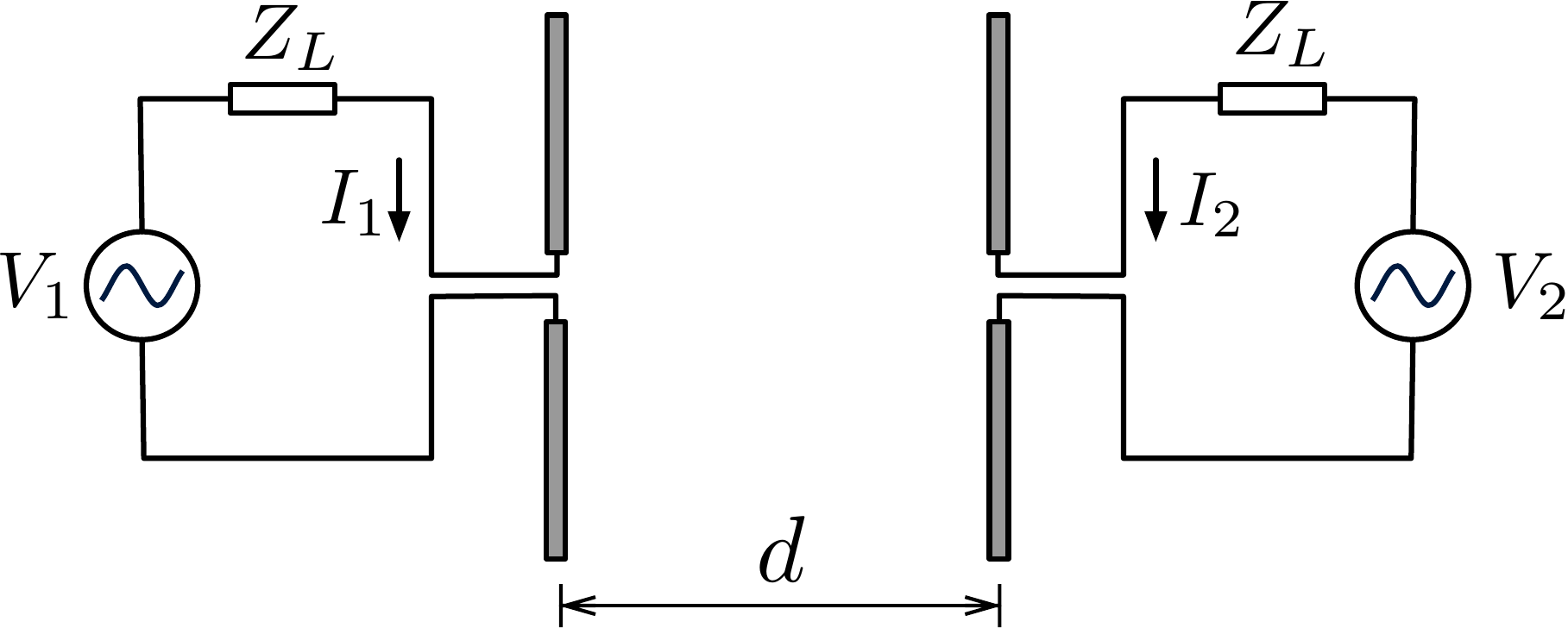}
\caption{The electric circuit for two active antennas used for near-field analysis  \label{f:circuit}}
\end{figure}

Assuming we have $n$ different transmitters, we get a $n \times n$ complex quadratic matrix with the impedance of a single dipole $Z_{nn}$
 in the diagonal.
For the analysis of multiple antennas, we follow the standard model and model each antenna \(i \in \{1, \ldots, n\}\) as an electric circuit with supply voltages \(\vec{V} = (V_1, \ldots, V_n)\), a resistor \(Z_L\), and the Hertzian dipole antenna, as shown in Fig.~\ref{f:circuit}. The current \(\vec{I} = (I_1, \ldots, I_n)\) at antennas \(1, \ldots, n\) can be calculated by

\begin{equation}
(Z_L \cdot \mathbf{I}_n + \mathbf{Z}) \vec{I} = \vec{V} \ , \label{zwoelf}
\end{equation}

where \( \mathbf{I}_n\) is the \(n \times n\) unity matrix (not to be confused with the current vector \( \vec{I}\)) and \(\mathbf{Z}\) the \(n \times n\) impedance matrix is defined as \(Z_{ii} = Z_{nn}\), and \(Z_{i,k}\) is the relative impedance using the distance \(d_{i,k}\) between antennas \(i\) and \(k\).

We define the {\bf normalized impedance matrix} as
\begin{equation}
\mathbf{M} = \frac{1}{Z_{nn}+Z_L} ( Z_L \cdot \mathbf{I}_n +\mathbf{Z}) \ . 
\end{equation}
Note that the linear equation system (\ref{zwoelf}) can be solved if and only if the determinant of the normalized impedance matrix is not zero. While this approach is  the standard in the community, to our knowledge the question whether this equation system is solvable for all antenna configurations has never been addressed, which will be the focus of this paper.
\subsection{Contributions}
We investigate two approximations of the Hertzian relative impedance. The far-field model  $f_{\mbox{\scriptsize far}}$ considers only the dominating term for $x\rightarrow \infty$, while The second approximation $f_{\mbox{\scriptsize mid}}$ takes the middle term into account.
\begin{equation}
	f_{\mbox{\scriptsize far}}(x) = \frac{3}{2}   \frac{\imaginary e^{- \imaginary x}}{x}\ ,
\end{equation}
\begin{equation}
	f_{\mbox{\scriptsize mid}}(x) = \frac{3}{2} \left(\imaginary \frac{e^{-\imaginary x}}{x}+ \frac{e^{- \imaginary x}}{x^2}\right) \ .
\end{equation}

We prove for the far-field approximation $f_{\mbox{\scriptsize far}}$ that already three antennas may be placed such that the corresponding impedance matrix has a determinant of $0$ for $Z_L =0$. 
For the more accurate approximation $f_{\mbox{\scriptsize mid}}$ we show that 15 antennas in a triangular configuration with a minimum distance of $\approx  0.65$ wavelengths  have a normalized determinant of $0$. 

For the Hertzian model the determinant of the normalized impedance model for a grid placement of $n\times n$ antennas with grid distance $\approx 0.65$  wavelengths strongly decreases for $Z_L=0$.
All results indicate that the current approach for determining the currents of near-field antennas should be used with great care.


\section{Unsolvable Antenna Configurations}

\subsection{Safe Configurations}
We start our investigation with the edge case, where antennas are placed very far apart.
Note that for $x \rightarrow \infty$ the function $f(x)$ as well as its approximations $f_{\mbox{\scriptsize mid}}$  and $f_{\mbox{\scriptsize far}}$ converge to $0$. Hence, all of the corresponding normalized  impedance matrices converge towards the identity matrix $\mathbf{I}_n$,  if all antenna distances are increased, since the diagonal entries are normalized to $1$.
Therefore, the determinant of the normalized impedance matrix converges to \(1\).

This observation can be strengthened by considering complex diagonally dominated matrices $\mathbf{A}$, where $|A_{ii}| > \sum_{k \neq i} |A_{ik}|$. It has been shown \cite{kolotilina2003nonsingularity} that those matrices are non-singular, i.e. that they have a non-zero determinant. Since the normalized matrix \(\mathbf{M}\) has diagonal entries $1$ and $n-1$ other entries for  $n$ antennas, then if the absolute value of these entries are smaller than \(\frac{1}{n-1}\), the matrix is diagonally dominated and thus  the determinant will be non-zero.

Now for $x\geq 1$ one sees that $|f(x)| \leq |f_{\mbox{\scriptsize far}}(x)| = \frac32 \frac1x$. So for  $x > \frac23 (n-1)$ the determinant of the Hertzian model will be non-zero. Remember that  $x = kd = \frac{2\pi}\lambda d $ for the distance $d$ of two antennas. If all mutual antenna distances obey 
\begin{equation}d> \frac1{3 \pi}  (n-1) \lambda \ ,\end{equation} then the determinant of the Hertzian and far-field impedance model is non-zero. Therefore, only antenna configurations where some antennas have distances $d \leq \frac1{3 \pi}  (n-1) \lambda $ bear the risk of resulting in having a zero determinant for their relative impedance matrix.

For the middle approximation $f_{\mbox{\scriptsize mid}}$ a similar result holds, using $|f_{\mbox{\scriptsize mid}}(x)| \leq \sqrt2 |f_{\mbox{\scriptsize far}}(x)|$ for $x\geq 1$. Then, for antennas where all mutual distances observe  \begin{equation}d> \frac{\sqrt2}{3 \pi}  (n-1) \lambda \ ,\end{equation} the determinant of the normalized impedance matrix using $f_{\mbox{\scriptsize mid}}$ is non zero.

\subsection{Far-Field Approximation}

\begin{figure}
\centering
\includegraphics[width=0.9\linewidth]{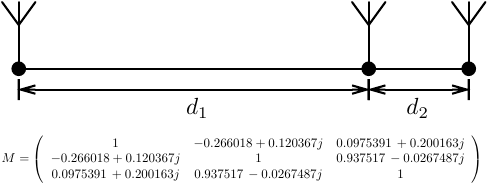}
\caption{Three antennas on a line with distances \( d_1 \) and \( d_2 \) \label{f:line}}
\end{figure}

We now consider three antennas on the line with distances \( d_1 \) and \( d_2 \) as shown in Fig.~\ref{f:line}
For the far-field function, we choose \( d_1 = 5.1373 \) and \( d_2 = 1.59932 \). It shows that for the normalized matrix \( [M] \) using \( f_{\text{\scriptsize far}} \), we get:  where 
\begin{equation*} [M] =  \small
\left(
\begin{array}{ccc}
 1 & -0.266018+0.120367 i & 0.0975391+0.200163 i \\
 -0.266018+0.120367 i & 1 & 0.937517-0.0267487 i \\
 0.0975391+0.200163 i & 0.937517-0.0267487 i & 1 \\
\end{array}
\right)
 \end{equation*}

where 
\begin{equation}
 |\det \mathbf{M} | \leq 4.5 \times 10^{-6}\ . \end{equation}

%

Clearly, this does not prove that the determinant is zero for these values. However, there is an elegant way to prove it. For this, we introduce a parameter \( t \in [0,1] \) and change the values \( x_1(t) \) and  \( x_2(t) \)  by moving it along the sinus curve:
\begin{equation} x_1(t) = 5.1373 + r \sin(2\pi t) \end{equation}
Similarly, we change \( d_2(t) \) with respect to  \( t \in [0,1] \) as:
\begin{equation} x_2(t) = 1.59932 + r \sin(2\pi(t-0.029)) \end{equation}

\begin{figure}
\centering
\includegraphics[width=0.99\linewidth]{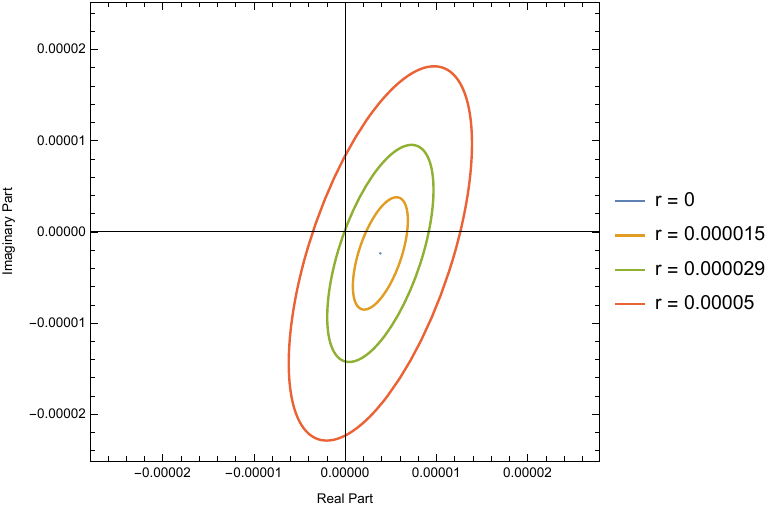}
\caption{The Jordan curves of \( \det(M) \) of the normalized impedance matrix using \( f_{\text{\scriptsize far}} \) for various \( r \) \label{f:jordanline}}
\end{figure}

Thus, for each \( t \), we obtain a normalized impedance matrix \(  \mathbf{M}_r(t) \), where each entry \(  \mathbf{M}_r(t)_{i,k} \) is a complex value depending on \( t \in [0,1] \). Since \( x_1(1) = x_1(0) \) and \( x_2(1) = x_2(0) \) , each value  in  \(  \mathbf{M}_r(t)_{i,k} \)  describes a continuous closed Jordan curve in \( \mathbb{C} \). By the definition of the determinant, the function \(\det( \mathbf{M}_r(t))\) also describes a closed Jordan curve from \( t = 0 \) to \( t = 1 \). Now, this Jordan curve encloses the origin \( 0 \) for \( r = 5 \times 10^{-5} \) as can be seen in Fig.~\ref{f:jordanline}. 

If  \( r \) slowly decreases to \( 0 \), this Jordan curve converges to a point  in a continuous way such that every point inside the Jordan curve is traversed. So, 
for every point inside the curve, there exists a value \( r' \) such that a curve contains this point. Since \( 0 \) is inside the a Jordan curve, there exists  a value \( r' \) for $r$  exactly hitting \(0\). By determining the corresponding value of \( t \), one may theoretically get the exact root of the determinant. Thus, this proves that there is a configuration of antennas on the line with distances $d_1$ and $d_2$ with \( k d_1= 5.1373 \pm 5 \times 10^{-5} \) and \( k d_2 = 1.59932 \pm 5 \times 10^{-5} \) such that \( \det \mathbf{M} = 0 \).

\begin{figure} \hfill
\includegraphics[width=0.3\linewidth]{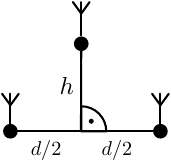} \hfill
\includegraphics[width=0.3 \linewidth]{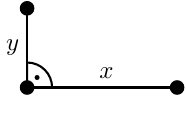} \hfill $ $ \\
\caption{Isosceles triangle and right angle antenna configuration \label{f:isoright}}
\end{figure}

Also for the isosceles configurations shown in Fig.~\ref{f:isoright} we obtain zero determinants for \( f_{\text{\scriptsize far}} \). For the isosceles triangle we the base length $d = 2.35477 \pm 0.00005$ and  height $h=1.25534 \pm 0.00001$ contains a determinant with value $0$. 
 For the right triangle we can choose $x= 2.07905 \pm 0.00001$ and $y=1.59907 \pm 0.00001$. 

Clearly, for $x \leq 3$ the function \( f_{\text{\scriptsize far}}(x)  \) does not approximate the function $f(x)$ very well and one may argue that these observations so far may be of some mathematical interest, but do not describe realistic antenna behavior. Therefore, we use this methodology and apply it to the function  \( f_{\text{\scriptsize mid}} \) and larger distances.

\subsection{Mid Field Approximation}

\begin{figure}\centering
\includegraphics[width=0.5\linewidth]{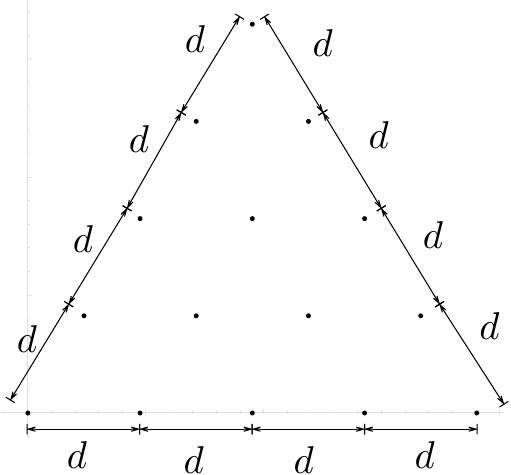}
\caption{Triangular placement of 15 antennas with distance $d$ \label{f:15triangle}}
\end{figure}

The function  \( f_{\text{\scriptsize mid}} \) approximates the Hertzian relative impedance much better than the far field function  \( f_{\text{\scriptsize mid}} \). It turns out that for three antennas using  \( f_{\text{\scriptsize mid}} \) the normalized impedance matrix is not singular for every antenna configuration. However, if the number of antennas increases, the situation changes. We have found the  triangular antenna configuration in Fig.~\ref{f:15triangle}, where a critical antenna position for $d = \frac1k 4.76$ is in the vicinity of $r=\frac1k 0.27$, i.e. if one can find a zero determinant position for antennas within a distance of $r$ of each antenna.

\begin{figure}
\centering
\includegraphics[width=0.99\linewidth]{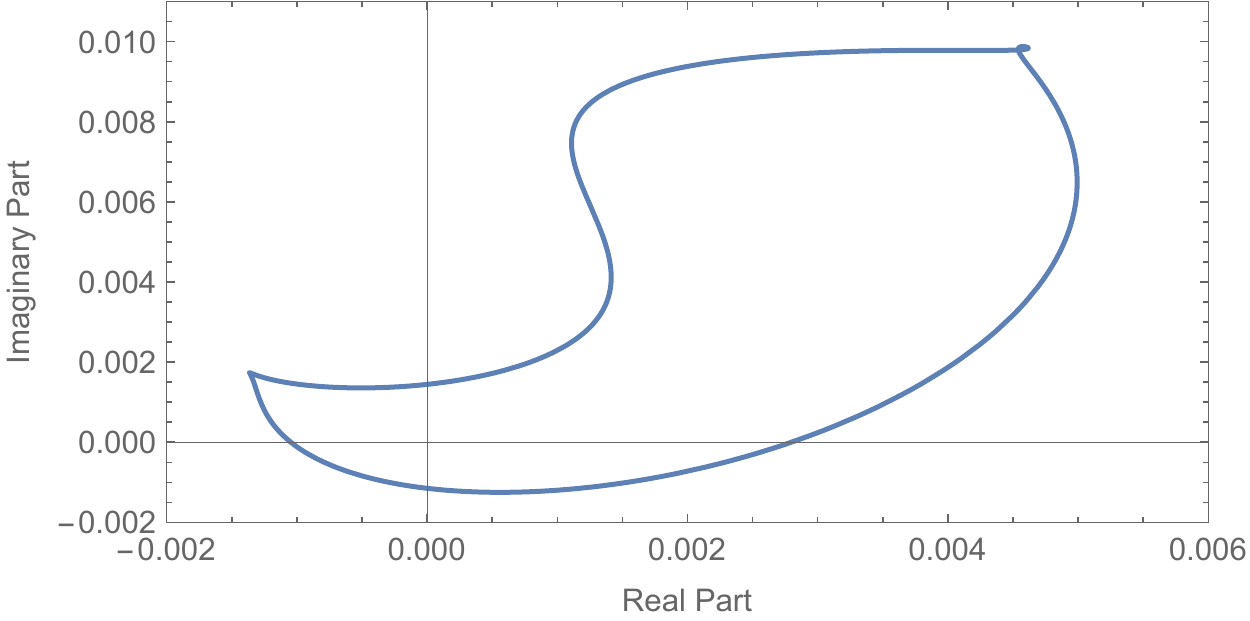}
\caption{The Jordan curves of \( \det(M) \) of the normalized impedance matrix using \( f_{\text{\scriptsize mid}} \) for 15 antennas \label{f:midtwitter}}
\end{figure}

In order to prove this statement, we consider a center point $p^0_i \in \mathbb{R}^2$ for each of the 15 antennas, where $i = \{1, \ldots, 15\}$. We define a curve for $t \in [0,1]$ with the function:

\begin{equation}
p_i(t) = p^0_i + (r \sin(2 \pi (t-\phi_i)), r \cos(2 \pi (t-\phi_i)))\ .
\end{equation}
This function defines the position of each antenna as a function of $t$, with $p^0_i$ as the center point and $r$ as the radius. We have found parameters $\phi_i$ such that 
the resulting Jordan curve of the determinant of the normalized impedance matrix encloses 0, as shown in Fig.~\ref{f:midtwitter}. Clearly, $0$ is enclosed by the Jordan curve given by $\det(M(t))$ und thus, a configuration with 15 antennas with zero determinant exists.

%
%
%
\begin{table}[htbp]
\centering
\caption{Antenna Placement of 15 Antennas}
\label{tab:antenna_params}
\begin{tabular}{cccc}
\toprule
$i$ & $x$ coordinate of $p^0_i$ & $y$ coordinate of $p^0_i$ & $\phi_i$ \\
\midrule

1 & 0 & 0 & 0.135353 \\
2 & 2.38 & 4.12228 & 1.24221 \\
3 & 4.76 & 8.24456 & 0.249188 \\
4 & 7.14 & 12.3668 & 0.464789 \\
5 & 9.52 & 16.4891 & 0.581601 \\
6 & 4.76 & 0 & 0.754519 \\
7 & 7.14 & 4.12228 & 1.28072 \\
8 & 9.52 & 8.24456 & 1.33471 \\
9 & 11.9 & 12.3668 & 0.517862 \\
10 & 9.52 & 0 & 1.32011 \\
11 & 11.9 & 4.12228 & 0.32972 \\
12 & 14.28 & 8.24456 & 0.56559 \\
13 & 14.28 & 0 & 1.06079 \\
14 & 16.66 & 4.12228 & 0.753963 \\
15 & 19.04 & 0 & 1.02783 \\
\bottomrule
\end{tabular}
\end{table}

%
%
%
%
%
%
%
%

%
%
%
The coordinates given are only approximation of the positions of the zero determinant case. Yet, we are able to prove (mathematically) its existence. For this we consider for each antenna a circular trajectory for a parameter $t \in[0,1]$ such that each antenna is rotated around the given position with some $\epsilon>0$ and and rotational offset. For each of the locations depending on $t$ we get a determinant which now also depends on $t$. Since, all relative distances are non-zero, the determinant of the normalized impedance is a continuous function $D(t)$ with respect to $t$ where $D(0)=D(1)$. So, $D(t)$ describes a closed Jordan Curve in $\mathbb{C}$.

As the simulations show, the Jordan Curve area encircle the origin $0$ of the complex plane. Thus, the inverse of impedance matrix does not exist.
\section{Grid Positioned Antennas under the Hertzian Model}
A special form of ULA antenna configurations is the grid, intensively used in 5G MIMO \cite{cui2022near}.
\begin{figure}[h]
	\centering
	\includegraphics[width=0.6\linewidth]{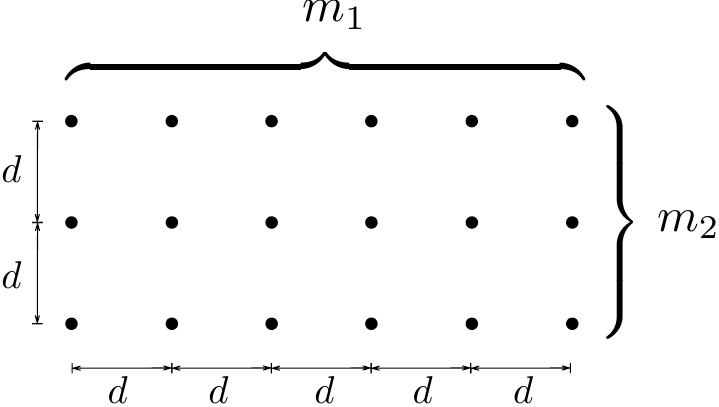}
	\caption{An $m_1 \times m_2$ grid configuration of $n=m_1 m_2$ antennas with distance $d$  \label{f:grid}}
\end{figure}

\begin{figure}[h]
	\centering
	\includegraphics[width=0.99\linewidth]{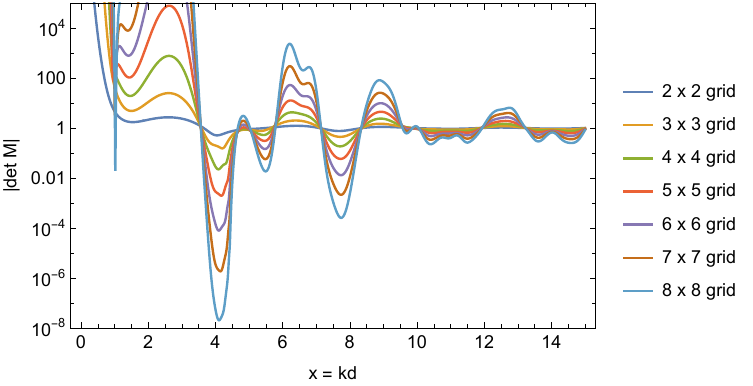}
	\caption{The absolute size of the determinant of the normalized matrix for the Hertzian model of a $m\times m$ grid  \label{f:grid-det}}
\end{figure}

For the exact Hertzian impedance model, we have not found an antenna configuration with determinant $0$.
As we have discussed above,  for  $d> \frac1{3 \pi}  (n-1) \lambda $ no zero determinant will occur. So, it seems intuitive that the risk for zero determinant grows with the number of antennas $n$. For this, we consider the grid placement of $n= m_1 \times m_2$ antennas in Fig.~\ref{f:grid}, where $n$ antennas are placed on a grid with distance~$d$.  Fig.~\ref{f:grid-det} shows that the absolute value of the determinant with respect to $d$. 

For the Hertzian model the determinant of the normalized impedance model for a grid placement of $n\times n$ antennas with grid distance $\approx 4.1/(2\pi)= 0.65$  wavelengths strongly decreases for $Z_L=0$.

\begin{figure}
	\centering
	\includegraphics[width=0.99\linewidth]{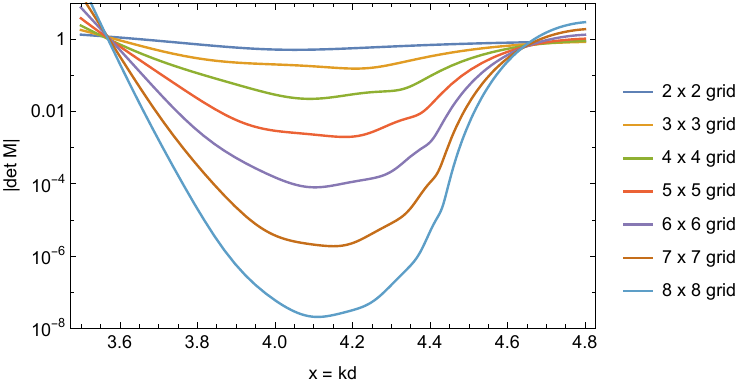}
	\caption{The relevant interval for the smallest determinant of the normalized matrix for the Hertzian model of a $m\times m$ grid  \label{f:grid-det45}}
\end{figure}

We see that absolute minima exist at $x \in[4.0,4.2]$ at different positions for each of the curves, see Fig.~\ref{f:grid-det45}, which shows that at this considerably larger antenna distance the solvability of the near-field is at stake. The minimum is around $x\approx 4.1$ and this translates to a minimum antenna distance of 
$ d \approx \frac{4.1}{2 \pi}  \lambda = 0.65 \lambda$ in the grid. Note that because of the existence of local minima in Fig.~\ref{f:grid} also for larger distances the absolute value decreases exponentially with the number of antennas in a grid.

\section{Conclusions and Outlook}
We have found different configurations of different isotropic antenna arrays, i.e. line, triangle, honeycomb, where the approximations $f_{\mbox{\scriptsize far}}$ and  $f_{\mbox{\scriptsize mid}}$  of the Hertzian impedance matrix form singular matrices by showing that the determinant is 0.  In order to prove the existence of the zero determinant, we use Jordan curves to prove the existence of the root of a complex function.
%

For the exact Hertzian model, the absolute value of the determinant of the impedance matrix decreases  for the square grid configuration with growing number of antennas. Thus, the existence of an multi antenna configuration with a determinant with value 0 cannot be ruled out and seems likely. 

If the determinant of the normalized impedance matrix is zero, then there is at the moment no method known to determine the (induced) currents at the antennas given the input voltage. 

We show that this circuit based approach is viable, if the mutual antenna distances are large enough, i.e. $d> \frac1{3 \pi}  (n-1) \lambda$ for $n$ antennas and wavelength $\lambda$. However, for smaller distances used for the near-field analysis, there is no result guaranteeing the solvability of the equation system to our knowledge. This is clearly a lack of understanding of the circuit based relative impedance approach for the near field and should be the focus of further research in the community.

The difficulties of this approach are shown for  approximations which are not far from the theoretical model, if one compares it to real-life impedances, which also deviate from the theoretical model. Thus, the solvability of the equations is not guaranteed there as well.

One explanation of this phenomenon may be that
by the conservation of energy, the radiated power can only be the feed-in power reduced by the power converted into work on the antenna and the Ohm losses.
The self-impedance in this model does not reflect the environment and its absolute size may be underestimated. 
Following this observation, a modification of the impedance matrix entries may be necessary and  may lead to equations systems, which are always solvable.\\
It remains to be examined whether this behavior is unique to the isotropic and Hertzian dipoles or whether it can  also be observed for other antenna models like e.g.  $\lambda/2$ 
dipoles \cite{zuhrt1953energieverhaltnisse}.
	\bibliographystyle{abbrv}
		\bibliography{nearfield-29}
	\newpage	

\end{document}